\documentstyle[12pt]{article}
\newcommand{\beq}{\begin{equation}}
\newcommand{\eeq}{\end{equation}}

\def\p1half{{\textstyle{{{p+1}\over{2}}}}}

\def\23phalf{{\textstyle{{{23-p}\over{2}}}}}

\begin{document}
\thispagestyle{empty}
\begin{titlepage}

\bigskip
\hskip 3.7in{\vbox{\baselineskip12pt
}}

\bigskip\bigskip
\centerline{\large\bf Duality Phase Transition in Type I String Theory}

\bigskip\bigskip
\bigskip\bigskip
\centerline{\bf Shyamoli Chaudhuri \footnote{Email:
shyamolic@yahoo.com} } \centerline{1312 Oak Drive}
\centerline{Blacksburg, VA 24060}
\date{\today}

\bigskip
\begin{abstract}
\vskip 0.1in We show that the duality phase transition in the unoriented
type I theory of open and closed strings is {\em first order}. The order parameter is the 
semiclassical approximation to the heavy quark-antiquark potential at finite temperature, 
extracted from the
covariant off-shell string amplitude with Wilson loop boundaries wrapped around the 
Euclidean time direction. Remarkably,
precise calculations can be carried out on either side of the phase boundary
at the string scale $T_C$$=$$1/2\pi \alpha^{\prime 1/2}$ by utilizing the T-dual, type IB
and type I$^{\prime}$, descriptions of the short string gas of massless gluon
radiation. We will calculate the change in the duality transition temperature in the 
presence of an electromagnetic background field.
\end{abstract}
\end{titlepage}

\section{Introduction}

\vskip 0.1in The canonical ensembles of the T-dual type IB, and type I$^{\prime}$, unoriented 
open and closed string theories are especially interesting because of the ease with
which the finite temperature Yang-Mills gauge theory limits can be extracted from the 
low energy approximation to the full string amplitude. 
Unlike the heterotic closed string theories, where the low energy limit yields a Yang-Mills
gauge theory coupled to a supergravity, here we can identify a self-consistent limit of any
perturbative string scattering amplitude which isolates pure nonabelian 
gauge theory physics. In this note, we will see how this observation can be applied
to infer the temperature dependence of the semiclassical heavy quark-antiquark potential
in the finite temperature nonabelian gauge theory.
We will find clear evidence of a {\em first order} phase transition at the string scale, 
inferring the precise transition temperature in
both the empty, flat target space background, as well as in the presence of a background electromagnetic
field strength. 

\vskip 0.1in 
In a previous work \cite{canon}, we have shown that the $T^2$ high
temperature growth of the string free energy at temperatures above the string scale
is compatible with a much more rapid $T^{10}$ growth at low temperatures,
exactly as would be expected in a finite temperature nonabelian gauge
theory. The latter result follows from isolating the leading contribution to
the free energy in the string mass level expansion, having performed the one-loop
modular integrals explicitly for this single term \cite{canon}. Could the fact that the 
free energy grows much more {\em slowly} at high temperatures far beyond the string 
scale be an indication of a thermal phase transition in the string ensemble, akin to the 
deconfining transition familiar from finite temperature gauge theory? And how could
one hope to extract evidence for a {\em first order} phase transition from perturbative string
theory, give that the amplitudes in any renormalizable and ultraviolet finite background 
are known to be analytic in their dependence on any moduli
parameters, in this case, the inverse temperature $\beta$. The key hint in favor of the 
possibility of evidence for such a 
phase transition is the
existence of distinct, T-dual string descriptions, type IB and type I$^{\prime}$,
on either side of the phase boundary at $T_C$. Thus, a matching of 
their respective low-energy 
gauge theory limits at the phase boundary can indeed display a discontinuity
 of first order, in an 
appropriately chosen physical observable. The observable of interest will turn
out to be the semi-classical heavy quark-antiquark potential.

\vskip 0.1in Our strategy for exposing such a phase
transition is as follows. Since the string one-loop vacuum amplitude has
shown no sign of a discontinuity, or non-analyticity, as a
function of temperature we must look at a different amplitude 
as plausible order parameter for the thermal
phase transition. A natural choice suggested by the correspondence
in the low energy limit to finite temperature Yang-Mills gauge
theory, would be the string theory analog of the expectation value
of a timelike Wilson-Polyakov-Susskind loop wrapping the Euclidean
time direction. Namely, the change in the string free energy in the
thermal vacuum due to the introduction of an external heavy quark,
generally taken to be the order parameter for the deconfinement
phase transition in finite temperature gauge theory
\cite{svet}. The expectation value of a single Wilson loop is related
to the disk amplitude in string theories 
\cite{alv}. We should note that this quantity is extremely
sensitive to the infrared divergences of finite temperature gauge
theory, necessitating ingenious techniques for a clear-cut study
of the order parameter in both the lattice, or dual confinement
model, approaches. It is generally considered easier
in nonabelian gauge theory to extract the desired result from a
computation of the {\em pair correlator} of Polyakov-Susskind
loops which yields the static heavy quark-antiquark potential in
the thermal vacuum. This is a fortunate circumstance, since the corresponding
string theory computation is readily accessible.

\vskip 0.1in In string theory, it turns out that the Polyakov path
integral summing surfaces with the topology of an annulus and with
boundaries mapped to a pair of fixed curves, ${\cal C}_1$, ${\cal
C}_2$, in the embedding target spacetime, wrapping the Euclidean
time coordinate, and with fixed spatial separation, $R$, can
be computed from first principles using Riemann surface
methodology \cite{cmnp}, and is unambiguously normalized unlike the
disk amplitude \cite{polchinskibook}. The requisite analysis is an interesting
extension of 
Polchinski's well-known covariant, one-loop vacuum amplitude computation
based on the Polyakov path integral \cite{poltorus,dbrane,polchinskibook}. 
The amplitude we will 
compute can be 
interpreted as an off-shell
closed string tree propagator, and the result in closed bosonic
string theory, but only in the limit that the macroscopic
boundaries, ${\cal C}_1$, ${\cal C}_2$, were point-like, was first
obtained by Cohen, Moore, Nelson, and Polchinski in 1986 \cite{cmnp}.
Their bosonic string theory analysis was adapted by myself, in
collaboration with Yujun Chen and Eric Novak in \cite{pair}, to 
address the limit of {\em large} macroscopic
loop length of interest here. The extension to the
macroscopic loop amplitude of the type I and type II superstring
theories with Dbranes by Novak and myself appears in \cite{pairf,flux}.
It should be noted that further extension to generic Wilson loop
computations should be possible, and our results by no means 
the end of this fascinating application to nonabelian gauge theory.

\vskip 0.1in We will calculate the pair
correlator of a pair of Polyakov-Susskind loops wrapping the
Euclidean time coordinate, extracting the low energy gauge theory
limit of the resulting expression where the contribution from
massive string modes has been suppressed. Notice that in the limit
of vanishing spatial separation, $R$ $\to$ $0$, the amplitude will
be dominated by the shortest open strings, namely, the gauge
theory modes in the massless open string spectrum. We can analyze 
this limit of the expression for its
dependence on temperature. We find clear evidence for a thermal
phase transition in the gauge theory at the
self-dual temperature: $T_{\rm C}$ $=$ $1/2\pi\alpha^{\prime
1/2}$. The temperature dependence in the limit of
vanishing spatial separation transitions from a $O(1/T^2)$ fall at low
temperatures to an $O(T^2)$ growth at high temperatures above the
string scale.
The argument is as follows. A Euclidean T-duality
transformation on our expression for the macroscopic loop
amplitude in the type IB string theory conveniently maps it to an
expression for a corresponding amplitude in the type I$^{\prime}$
string theory. That expression will be well-defined in the
temperature regime above $T_C$, or vice versa,
and the low energy gauge theory
limit can be easily taken as before. Thus, the existence of T-dual
type IB, and type I$^{\prime}$, descriptions of the thermal ground
state enable precise computations to be made in the low energy
gauge theory limit on either side of the phase boundary at $T_C$. 

\vskip 0.1in The
intuition that a gas of short open strings transitions into a high
temperature long string phase is an old piece of string folklore, and could
find important application in both gauge theory and early universe
cosmology, including the physics of cosmic strings \cite{long}. However,
as is clear from the detailed analysis given by us in \cite{canon}, despite the
exponential growth in the degeneracies of states with high mass level
number in the string spectrum, the
free energy itself is free of both divergences, and of non-analyticity!
In particular, it is not true that the superstring canonical ensemble
 \lq\lq breaks down" above a string scale  \lq\lq limiting" temperature; the
canonical ensemble is well-defined on both sides of the phase 
boundary at $T_C$. Rather, it is necessary to carry out a T-duality 
transformation to probe the novel high temperature behavior.
  
\section{Evidence for the Long String Phase Transition}

\vskip 0.1in A plausible order parameter signalling a thermal
phase transition in the type IB string theory at a temperature of
order the string mass scale is suggested by the correspondence
with the low energy finite temperature gauge theory limit. It is
well known that the order parameter signalling the thermal
deconfinement phase transition in a nonabelian gauge theory at
high temperatures is the expectation value of a closed timelike
Wilson-Polyakov-Susskind loop \cite{svet}. We
wish to investigate evidence for a thermal 
phase transition in the low energy gauge theory limit of
type I string theory at a temperature of order
the string scale. Such a phase transition has long been
 conjectured to arise in the massless radiation 
gas of short open strings, characterized plausibly by 
long string formation
in the high temperature phase \cite{long}, and often interchangeably
referred to as a \lq\lq Hagedorn" phase transition.

\vskip 0.1in Since the one-loop free energy in the type
IB thermal vacuum displays no non-analyticity, or discontinuities,
as a function of temperature as shown by us in \cite{canon}, it is natural to look for evidence
in a different string amplitude. A natural choice suggested by the
correspondence in the low energy limit to finite temperature
Yang-Mills gauge theory, would be the string theory analog of the
expectation value of a timelike Wilson loop wrapping the Euclidean
time direction. Namely, the change in the free energy in the
thermal vacuum due to the introduction of an external heavy quark,
generally taken to be the order parameter for the deconfinement
phase transition in finite temperature gauge theory, pointed
out, independently, by both Polyakov and Susskind \cite{svet}. 

\vskip 0.1in As mentioned in the introduction, a convenient
starting point is the Polyakov path integral summing surfaces with
the topology of an annulus and with boundaries mapped to a pair of
fixed curves, ${\cal C}_1$, ${\cal C}_2$, in the embedding target
spacetime, wrapping the Euclidean time coordinate, and with fixed
spatial separation, $R$. This macroscopic loop amplitude
can also be computed from first
principles using Riemann surface methodology, an extension of the
covariant one-loop string vacuum amplitude derived by Polchinski in
\cite{poltorus,polchinskibook,dbrane}. The macroscopic loop
amplitude is an off-shell
closed string tree propagator, and the result in closed bosonic
string theory, but only in the limit that the macroscopic
boundaries, ${\cal C}_1$, ${\cal C}_2$, were point-like, was first
obtained by Cohen, Moore, Nelson, and Polchinski \cite{cmnp}, 
and
extended to include the limit of {\em large} macroscopic loop
lengths of interest here by myself in collaboration with Yujun
Chen and Eric Novak in \cite{pair}. The low energy gauge theory limit of the
macroscopic amplitude yields the pair
correlator of Wilson loops wrapping the Euclidean time
coordinate in the finite temperature nonabelian 
gauge theory living on the worldvolume of D9branes, 
with the contribution from massive string
modes suppressed. Notice that in the limit of vanishing
spatial separation, $R$ $\to$ $0$, the target spacetime bosonic
contribution to the string one-loop amplitude will be dominated
by the shortest (bosonic) open strings, namely, the gauge theory modes in
the massless open string spectrum. We will analyze
this field theoretic
limit of the macroscopic loop amplitude for its dependence on temperature.

\vskip 0.1in Consider, therefore, the pair correlation function of a pair of
Polyakov-Susskind loops lying within the worldvolume of the
D9branes, and with fixed spatial separation $R$ in a direction
transverse to compactified Euclidean time, $X^0$. Recall that the
boundaries of the worldsheet are the closed \lq\lq
world-histories" of the open string endpoint, which couples to the
gauge fields living on the worldvolume of the Dbranes. The
endpoint states are in the fundamental, and anti-fundamental, representations of the gauge
group. Thus, when the closed worldlines are constrained to coincide
with closed timelike loops in the embedding spacetime, the
resulting string amplitude has a precise correspondence in the low
energy gauge theory limit to the correlation function of two closed timelike
loops representing the spacetime histories of a static,
heavy \lq\lq quark--antiquark" pair with fixed spatial separation. Since we
wish to probe the high temperature behavior of the low energy
gauge theory limit, we should use the Euclidean T-dual type
I$^{\prime}$ description of the thermal vacuum.
The result for the pair correlator of temporal Wilson
loops in the Euclidean T-dual type I$^{\prime}$ vacuum, ${\cal
W}^{(2)}_{\rm I^{\prime}}$, derived from first principles from an
extension of the ordinary Polyakov path integral in the references
\cite{cmnp,pair}, takes the remarkably simple form
\cite{decon}:
\begin{eqnarray}
{\cal W}^{(2)}_{\rm I^{\prime}} (R,\beta)  =&& \lim_{\beta \to 0}
\int_0^{\infty} dt {{e^{- R^2 t/2\pi \alpha^{\prime}
}}\over{\eta(it)^{8}}}
       \sum_{n \in {\rm Z}} q^{n^2 \beta^2 /4 \pi^2 \alpha^{\prime} } \cr \quad 
        && \times \left [
 ({{\Theta_{00}(it;0)}\over{\eta(it)}})^4
 -  ({{\Theta_{01}(it;0)}\over{\eta(it)}})^4    \right ] 
\label{eq:pairc}
\end{eqnarray}
The summation variable, $n$, labels closed string winding modes in
this expression,
each of which wraps around the Euclidean time-like coordinate
$X^{0\prime}$. The analyis in \cite{canon} includes both half integer
and integer moding in the thermal winding spectrum, but let us restrict ourselves
to just the integer modes in this calculation; the half-integer moded spectrum
makes a similar contribution. 

\vskip 0.1in We need to identify a suitable physical observable that can be
extracted from this calculation. 
An observable of considerable significance in nonabelian gauge
theory is the
semiclassical heavy quark-antiquark potential.
The low energy finite temperature gauge theory limit
of the macroscopic loop amplitude yields the temperature
dependence of the static heavy quark potential as follows. Let us
set:
\begin{equation}
{\cal W}^{(2)}_{\rm I^{\prime}} (\beta)  = \lim_{s \to\infty}
\int_{-s}^{+s} d s ~ V[R,\beta] \quad , \label{eq:poten}
\end{equation}
where $s$ is the proper time parameterizing the worldlines of
the quark-antiquark pair, wrapping the Euclidean time coordinate.
We can invert this relation to express $V[R,\beta]$ as an integral
over the modular parameter $t$ \cite{dkps,polchinskibook}.
Consider the $q$ expansion of the integrand valid for
$t$$\to$$\infty$, where the shortest open strings dominate the
modular integral. The low temperature behavior is best extracted
from the type I$^{\prime}$ amplitude, with its spectrum of thermal
winding modes. Retaining the leading terms in the $q$ expansion
and performing explicit term-by-term integration over the
world-sheet modulus, $t$, isolates the following interaction
potential at temperatures below the string mass scale
\cite{decon,pair}:
\begin{eqnarray}
V(R,\beta)|_{\beta >> \beta_C } =&&  (8\pi^2
\alpha^{\prime})^{-1/2} \int_0^{\infty} dt e^{- R^2 t/2\pi
\alpha^{\prime} } t^{1/2}
 \sum_{n =-\infty }^{\infty} \left [ 16 + O(e^{-\pi t}) \right ]  q^{n^2 \beta^2
/4 \pi^2 \alpha^{\prime} } \cr \quad \simeq &&(8\pi^2
\alpha^{\prime})^{-1/2} \Gamma (3/2)
 (2\pi \alpha^{\prime} )^{3/2}  {{2^4}\over{R^3}} \left [ 1 -{{3}\over{2}}  \sum_{n=-\infty}^{\infty}
 {{\beta^2 n^2 }\over{R^2}} \right ]   ~ \cr
\equiv &&
   16 \pi^{1/2} \alpha^{\prime} \Gamma (3/2) {{1}\over{R^3}}
 \left [ 1   - 3 ~ \zeta (-2 , 0 ){{\beta^2}\over{R^2}} \right ] \quad .
\label{eq:static}
\end{eqnarray}
Note that at low temperatures, 
$\beta$ $>>$ $\beta_C$, the power series expansion 
gives an $O(\beta^2 /R^5)$ correction to the zero temperature
static potential. Here, $\beta_C$ is the inverse
self-dual temperature. Above $T_C$, we can extract the correct form
of the potential from the low energy gauge theory limit of the 
T-dual type IB vacuum amplitude. The corrections to the zero temperature
static potential take the form of a power series in $(\beta^4_C /\beta^2 R^2)$
at high temperature:
\begin{eqnarray}
V(R,\beta)|_{\beta << \beta_C } =&&  (8\pi^2
\alpha^{\prime})^{-1/2} \int_0^{\infty} dt e^{- R^2 t/2\pi
\alpha^{\prime} } t^{1/2} \cr \quad &&\quad \times \sum_{n = - \infty}^{\infty}
\left [  16   ~+~ O(e^{-\pi t}) \right ] q^{4
\pi^2 \alpha^{\prime} n^2 /\beta^2} + \cdots \cr \quad \simeq&&
(8\pi^2 \alpha^{\prime})^{-1/2} \Gamma (3/2)
 (2\pi \alpha^{\prime} )^{3/2}  {{2^{4}}\over{R^3}} \left [ 1 - {{3}\over{2}} \sum_{n=-\infty}^{\infty}
 {{(4\pi^2 \alpha^{\prime })^2  n^2
}\over{\beta^2 R^2}} \right ]  \cr \equiv &&
16 \pi^{1/2} \alpha^{\prime}
 \Gamma (3/2) {{1}\over{R^3}}
\left [ 1 - 3 \zeta (-2, 0 ) {{ \beta_C^4 }\over{\beta^2
R^2}} \right ] \quad , \label{eq:statici}
\end{eqnarray}
Notice that the discontinuity is in
the {\em first derivative} with respect to temperature. Remarkably, 
precise computations can
nevertheless be carried out on either side of the phase boundary
by utilizing, respectively, the low energy gauge theory limits of
the {\em pair} of thermal dual string theories, type IB and type
I$^{\prime}$. The result clarifies that the thermal phase transition
at $T_C$ $=$ $1/2\pi\alpha^{\prime 1/2}$ in the gauge theory
is {\em first order}.

\vskip 0.1in Finally, it is helpful to note that in the presence of a background
electromagnetic field, the result for the transition temperature is simply
altered to $T_D$ $=$ $1/ 2\pi \alpha^{\prime 1/2} u$, where ${\cal F}_{9j}$ $=$ 
${\rm tanh}^{-1} u$ is the constant part of the magnetic field strength, and $j$ labels
any other spatial coordinate. This fact was
noted by us in \cite{decon,coup}, but without appreciation of the full significance of
the thermal duality transition.

\section{Conclusions}

\vskip 0.1in We have shown that the duality phase transition in the unoriented
type I theory of open and closed strings is {\em first order}. The order parameter is the 
semiclassical approximation to the heavy quark-antiquark potential at finite temperature, 
extracted from the
covariant off-shell string amplitude with Wilson loop boundaries wrapped around the 
Euclidean time direction. Remarkably,
precise calculations can nevertheless be carried out on either side of the phase boundary
by utilizing the respective low energy limits of the T-dual, type IB
and type I$^{\prime}$, string theories. Either describes the short string gas of massless gluon
radiation, but on one, or other, side of the phase boundary. We will leave discussion
of the applications of this result in both gauge theory and cosmology \cite{long} to future work.

\vspace{0.1in}\noindent{\bf Acknowledgements:} I would like to thank David
Gross for a helpful discussion. This research has been supported in 
part by the NSF, the Aspen Center for Physics, 
and the Kavli Institute for Theoretical Physics.

\end{document}